\documentclass{iaus}
\usepackage{graphicx}

\def\vhel{\ifmmode{V_{{\rm HEL}}}\else{$V_{{\rm HEL}}$}\fi}
\def\vsys{\ifmmode{V_{\rm sys}}\else{$V_{\rm sys}$}\fi}
\def\kms{\ifmmode{~{\rm km\,s}^{-1}}\else{~km~s$^{-1}$}\fi}
\def\vlsr{\ifmmode{v_{\rm lsr}}\else{$v_{\rm lsr}$}\fi}

\title[short title of paper] 
{New Planerary Nebulae towards the Galactic bulge}

\author[P. Boumis et al.]   
{P. Boumis$^1$, S. Akras$^1$, P. A. M. van Hoof$^2$, G. C. Van de Steene$^2$, \break J. Papamastorakis$^3$ \and J. A. Lopez$^4$}

\affiliation{$^1$Institute of Astronomy \& Astrophysics, National Observatory of Athens, I. Metaxa \& V. Pavlou, GR-152 36 P. Penteli, Athens, Greece \break 
$^2$ Royal Observatory of Belgium, Ringlaan 3, B-1180 Brussels, Belgium \break 
$^3$ Department of Physics, University of Crete, GR-710 03 Heraklion, Crete, Greece \break 
$^4$ Instituto de Astronomía, UNAM, Apdo. Postal 877, Ensenada, BC 22800, Mexico \break 
}

\pubyear{2006}
\volume{234}  
\pagerange{1--2}
\date{?? and in revised form ??}
\jname{Planetary Nebulae in Our Galaxy and Beyond}
\editors{M. J. Barlow \& R. H. M\'endez, eds.}
\begin{document}

\maketitle

\begin{abstract}

New Planetary Nebulae (PNe) were discovered through an [O {\sc iii}]
5007 \AA~emission line survey in the Galactic bulge region with $l >$
0$^{\circ}$. We detected 240 objects, including 44 new PNe. Deep
H$\alpha +$[N {\sc ii}] CCD images as well as low resolution spectra
were obtained for the new PNe in order to study them in
detail. Preliminary photo--ionization models of the new PNe with
Cloudy resulted in first estimates of the physical parameters and
abundances. They are compared to the abundances of Galactic PNe.
\end{abstract}

\firstsection 
\section{Introduction}

Galactic Planetary Nebulae (PNe) are of great interest because of
their important role in the chemical enrichment history of the
interstellar medium as well as in the stellar evolution of our Galaxy
(\cite[Beaulieu {\it et al.} 2000]{Beaulieu00} and references therein). Many
surveys have been made in the past in order to discover new PNe
(\cite[Boumis {\it et al.} 2003]{Boumis03} and references therein -- Paper
I, \cite[Boumis {\it et al.} 2006]{Boumis06} -- Paper II, \cite[Parker {\it et
al.} 2006 and references therein]{Parker06}).

\section{Observations}\label{sec:obs}

The survey was performed during the 2000--2001 observing seasons with
the 0.3 m telescope at Skinakas Observatory in Crete, Greece. 
Our aim was to discover PNe which are extended or pointlike showing
strong [O {\sc iii}] 5007 \AA~emission with a signal-to-noise ratio
greater than 4. The observational details and the detection method
are given in Paper I and Boumis \& Papamastorakis (2001),
respectively. Follow--up observations (images and spectra) were
obtained with the 1.3 m telescope at the same site during 2001--2003
and complementary spectral observations during 2005--2006. The images
were obtained in H$\alpha +$[N {\sc ii}] in order to study
the morphology of the PNe and also measure their angular extent while
their low--resolution spectra confirmed their photo-ionized
nature. All new PNe can be seen in Papers I and II.

\begin{table}
\centering
\label{table1}
\begin{tabular}{lllllll}
\hline
{\em Parameters} &
\multicolumn{1}{c|}{PTB17} & \multicolumn{1}{c|}{PTB26} & \multicolumn{1}{c|}{PTB31} & \multicolumn{1}{c}{PTB34} &
\multicolumn{1}{c|}{mean bulge$^g$} & \multicolumn{1}{c|}{mean disk$^g$} \\
\hline
log(L/L$_\odot$)& 2.80  & 2.78  & 3.82  & 4.12  & & \\
T$_{eff^b}$   & 111.4 & 65.0  & 56.9  & 85.9  & & \\
n$_{e^d}$     & 0.17  & 0.06  & 0.18  & 0.59  & &\\
T$_{e^b}$     &  9.8  & 6.49  & 9.72  & 8.20  & &\\
log(M/$M_\odot$)& 0.19  & 0.46  & -0.16 & -0.61 & &\\
radius in$^e$   & 0.20  & 0.28  & 0.10  & 0.05  & &\\
radius out$^e$  & 0.40  & 0.72  & 0.29  & 0.21  & &\\
filling factor  &  1.00 & 1.00  &  1.00 &  0.58 & &\\
log(U)          & -2.75 & -2.53 & -1.11 & -0.72 & &\\
dust/gas$^a$    &  5.56 & 6.15  &  6.08 &  5.66 & &\\
distance$^f$    &  7.80 & 7.80  &  7.80 &  7.00 & & \\
$\epsilon$(He)    & 11.17 & 11.02 & 11.01 & 11.14 & 11.09 & 11.06\\
$\epsilon$(O)     &  8.47 & 9.16: &  8.35 &  9.06 &  8.66 &  8.67\\
$\epsilon$(N)     &  8.12 & 8.01  &  8.32 &  8.34 &  8.43 &  8.34\\
$\epsilon$(S)     &  6.58 & 6.79  &  7.10 &  7.55 &  7.05 &  6.93\\
$\epsilon$(Ne)    &  $-$  & 8.00$^\dag$  &  8.00$^\dag$ &  $-$  &  8.03 &  8.08\\
$\epsilon$(Ar)    &  $-$  & $-$   &  $-$  &  6.96 &  6.60 &  6.42\\
\hline
\end{tabular}
\caption[]{The physical parameters of our sample's PNe determined with
Cloudy. (a) 10$^{-3}$, (b) kK, (c) Log(X/H)$+$12, (d) 10$^3$ cm${-3}$,
(e) pc, (f) kpc, (g) Exter et al. (2004), ($\dagger$) fixed at an
assumed value, ($\colon$) the value is uncertain.}
\end{table}

\section{Preliminary Photoionization Results}\label{sec:photoion}

A number of different techniques are in use to determine the physical
parameters of PNe. In our case, we decided to use the photo-ionization
code Cloudy, last described by \cite[Ferland et al. (1998)]{Ferland98},
since this code is widely used and has been tested for many different
physical conditions. In this work, we present preliminary results for
four of our new PNe. Full results will be presented in \cite[Akras et
al. (2006)]{Akras06}. The model assumptions can be found in \cite[van
Hoof \& Van de Steene (1999)]{vanhoof99}. It should be noted that as
nebular distances we used both the fixed bulge distance (7.8 kpc), and
distances determined with the method described in \cite[Van de Steene
\& Zijlstra (1995)]{VandeSteene95}.
All results will be presented analytically in
\cite[Akras et al. (2006)]{Akras06}. The observed quantities used to
derive the physical parameters of our PNe are (a) the emission line
spectrum of each nebula, (b) their angular diameters and (c) the total
hydrogen flux. Further modeling will be performed including radio
and/or infrared data whenever they are available. The resulting
physical parameters for our sample PNe determined with Cloudy are
given in Table 1. We also present the mean bulge and disc abundances
taken from \cite[Exter et al. (2004)]{Exter04}. 

\begin{acknowledgments}

AS and PvH acknowledge financial support from
the Belgian Science Policy Office project IUAP P5/36.
Skinakas Observatory is a collaborative project of the
University of Crete, the Foundation for Research and
Technology--Hellas and the Max--Planck--Institut fur
Extraterrestrische Physik.
\end{acknowledgments}

\end{document}